# Physics Education for Capacity Development and Research in Africa

Sam Ramaila

*University of Johannesburg, South Africa*

**Abstract**

The acceleration of socio-economic development is intrinsically linked to the level of scientific development. While the existing scientific interventions within Africa played a transformative role in the enhancement of human capital development, adequate investment in research and development is required to make further significant strides going forward. Unlocking Africa's potential requires sustained investment in research and development. However, inadequate expenditure in research and development as a percentage of gross domestic product by African countries does not augur well for the progressive realization of sustainable scientific development. This contribution highlights challenges afflicting physics education in Africa and provides a reflection on key areas for intervention to strengthen capacity building. Critical interrogation of enablers and constraints is required in order to harness the efficacy of capacity building efforts with a view to engender fundamental transformative change in relation to meaningful enhancement of human capital development in Africa. Reconfiguration of the existing scientific interventions some of which yielded remarkable results remains a key strategic imperative in the long to medium term. Progressive realization of this key strategic imperative hinges to a large degree on the establishment of collaborative partnerships involving African key stakeholders. Contextually appropriate recommendations for coherent acceleration of scientific development within the broader African context are advanced.

## 1. Introduction

Capacity development and sustainable scientific development in Africa are plagued by existing pervasive fundamental challenges. These challenges include lack of requisite facilities to conduct research at many African institutions and universities, inadequate engagement by physicists in research and academia, low doctoral enrolment and graduates, low research publications as well as paucity of research institutes and industry which limits job opportunities for physicists to teaching in secondary schools and universities [1, 2]. There is a critical need to put appropriate scientific interventions in place which are essentially geared towards the improvement of the quality of physics education in Africa. In response to this key strategic imperative, the Association of Commonwealth Universities (ACU) and the Institute of Physics (IOP) created an evidence base for a potential multi-year programme to improve physics training, research, infrastructure and collaboration in nine countries in Sub Saharan Africa [3]. These countries are Ethiopia, Ghana, Kenya, Malawi, Nigeria, Rwanda, South Africa, Tanzania and Uganda. This enormous undertaking represents a significant step towards strengthening physics education for capacity development and research in Africa.

---

\*Corresponding Author

*Email address:* ramailasam8@gmail.com (Sam Ramaila)

## 2. Capacity development and research

Physics education provides a solid basis for progressive realization of capacity development and robust pursuit of scientific research in Africa. However, a range of interventions are required to strengthen human capacity development. Strengthening human capacity development should be predicated on clearly identified key areas for intervention. Key areas for intervention identified by the Association of Commonwealth Universities and Institute of Physics include gender inclusivity, training and education, academic and staff capacity, innovation and commercialization as well as collaboration and networks [3]. The under-representation of women in physics remains a major structural problem in Africa. Physics training and education ought to foster the inclusion of women as an integral part of capacity development. In addition, African universities need to create opportunities for establishing vibrant partnerships that promote collaborative construction of scientific knowledge. The advent of the Fourth Industrial Revolution provides enormous opportunities for African universities to embrace digital transformation with a view to foster innovation and commercialization.

## 3. Unlocking Africa's potential through physics education

Physics education has the potential to play a pivotal role in unlocking Africa's potential. Sustainable pursuit of scientific research in fundamental and applied physics requires highly skilled physicists. By its very nature, effective physics education can be harnessed as a vital means to train the next generation of physicists required to unlock Africa's potential. Other creative mechanisms identified to unlock Africa's potential include addressing Africa's inability to fill positions in physics fields, critical reflection on future aspirations and the role of physics education, and rethinking the role of teachers and other stakeholders in physics education [4]. Filling positions in physics fields in Africa remains a perennial challenge which stems from existing limited critical mass of available physicists. This perennial challenge stifles Africa's global scientific competitiveness.

## 4. Existing interventions promoting scientific development in Africa

There is a number of existing interventions promoting scientific development in Africa. The South African Institute of Physics (SAIP) coordinates a Teacher Development Project which is primarily aimed at teacher professional development in South Africa [5]. The project program is accredited by the South African Council for Educators (SACE). SACE is the professional council for educators that aims to enhance the status of the teaching profession through appropriate registration, management of professional development and inculcation of a Code of Ethics for all educators [6]. Teachers' participation in the SAIP Teacher Development Project enables them to accumulate Continuous Professional Development (CPD) Points which are essential for career enhancement. The National Astrophysics and Space Science Programme (NASSP) is another existing intervention that promotes scientific development in Africa. NASSP is a multi-institutional initiative funded by the Department of Science and Innovation through the National Research Foundation to train South African students in Astrophysics and Space Science at Honours and Master's levels and to provide a pipeline to PhD studies in these and related research areas [7]. At another pragmatic level, the African Union (AU) established Innovating Education in Africa Initiative in 2018. The initiative is aimed at identifying, promoting and supporting the systemic adoption and replication of education innovations in all aspects of education and training in Africa, fostering policy dialogues with policy makers and development stakeholders to make the case for embedding innovation in education systems in Africa and endorsing the implementation of the continental programme [8]. As a dynamic and vibrant scientific intervention, the African School of Fundamental Physics and Applications made significant strides in human capacity development in Africa since its inception [9]. This monumental achievement is evidenced by a substantial number of students from various African countries who benefited immensely through active participation in this scientific intervention. The success achieved through the implementation of the aforementioned scientific interventions underscores the need for more interventions to be put in place across the African continent with a view to unlock Africa's potential.

**5. Attraction and retention of students in Physics**

Concerted efforts are required to attract and retain students in Physics. Attraction and retention of students in Physics can be realized through adoption of appropriate and sustainable mechanisms. These mechanisms may include:

- Assessing the quality of students entering African universities.
- Rethinking student under-preparedness for higher education.
- Assessing the state of undergraduate physics teaching and learning in African universities.
- Developing a set of standards for Physics training in Africa.
- Assessing levels of commonality and diversity of the physics programmes in Africa.
- Assessing range, scope and effectiveness of current teaching and learning practices in Physics at African universities.
- Developing a set of contextually appropriate recommendations aimed at improving the effectiveness of Physics teaching at African universities.

**6. Recommendations**

Scientific development remains a key strategic imperative in Africa. Coherent realization of this imperative ought to be informed by contextually appropriate recommendations. The following proposed recommendations are advanced to inform scientific development within the broader African context:

- There is a need for enhanced coordination of existing scientific interventions.
- Regular assessment of the impact and efficacy of existing scientific interventions is imperative.
- There is a need to go beyond capacity building to track and monitor the progress made through coherent implementation of existing scientific interventions.

- Active involvement of African key stakeholders in scientific development is imperative.
- Enhanced coordination of African efforts aimed at acceleration of meaningful scientific development is absolutely essential.
- It is imperative for African universities to embrace evidenced-based physics education research that fosters adequate exposure to plurality of knowledge epistemologies.
- Increased investment in scientific development is crucial for unlocking Africa's potential.
- Better funding of African universities is a timely and necessary strategic intervention.
- There is a need to establish African scientific evaluation and monitoring committee.

## 7. Conclusion

Physics education is crucial for the acceleration of socio-economic development in Africa. Strengthening physics education remains a key requirement for capacity development and sustainable pursuit of scientific research within the broader African context. Existing scientific interventions ought to be sustained and augmented to ensure skills development through active involvement in physics education activities. African countries face the key imperative to address inadequate expenditure in research and development as a percentage of gross domestic product with a view to facilitate sustainable scientific development in its broadest sense.

## 8. References

[1] Sa'id, R.S., Fuwape, I., Dikandé, A.M. et al. Physics in Africa. Nature Reviews Physics, 2, 520–523 (2020). doi: https://doi.org/10.1038/s42254-020-0239-8.

[2] Department of Higher Education and Training. Statistics on Post-School Education and Training in South Africa (2017). doi: https://www.dhet.gov.za.

[3] The Association of Commonwealth Universities and Institute of Physics. Africa-UK Physics Partnership Programme Feasibility Study Report (2020). doi: https://www.acu.ac.uk/media/3533/feasibility-study-report-final.pdf.

[4] Sichangi, M. How science education can unlock Africa's potential (2018). doi: http://www.adeanet.org/en/blogs/how-science-educationcan-unlock-africa-s-potential.

[5] South African Institute of Physics. Physics Teacher Development Project. doi: http://oldsite.saip.org.za/index.php/projects/physics-teacher-development.

[6] South African Council for Educators. doi: https://www.sace.org.za.

[7] The National Astrophysics and Space Science Programme. doi: https://www.star.ac.za.

[8] African Union. Innovating Education in Africa Initiative (2018). doi: https://au.int/en/pressreleases/20181005/innovation-education-africa-expo-2018-kicked-today.

[9] African School of Fundamental Physics and Applications. doi: https://www.africanschoolofphysics.org.